\documentclass[12pt]{article}
\usepackage{epsfig}
\usepackage [small] {caption}
\newcommand{\ra}{\rightarrow}
\newcommand{\bq}{\begin{eqnarray}}
\newcommand{\eq}{\end{eqnarray}}
\newcommand{\ov}{\overline}

\begin{document}

\begin{center}{\boldmath {$\gamma\gamma\ra\pi\pi\,,\, KK :\,\,$}
{\bf leading term QCD vs handbag model}}\end{center}

\begin{center}{\bf Victor L. Chernyak}
\footnote{\,\,\, E-mail: v.l.chernyak @ inp.nsk.su}
\end{center}

\begin{center}
{Budker Institute of Nuclear Physics,\\
630090 Novosibirsk, Russia}
\end{center}
\vspace{0.5cm}
\begin{center}{\bf Abstract}\end{center}

The "handbag" model was proposed as an alternative, at the present day energies,
to the leading term QCD predictions for some hard exclusive processes. The recent
precise data from the Belle Collaboration on the large angle cross sections $\gamma\gamma
\ra\pi\pi\,,\, KK$ allow a check of these two approaches to be performed.

It is shown that the handbag model fails to describe the data from Belle,
while the leading term QCD predictions are in  reasonable agreement with these data.
\vspace{0.5cm}

\begin{center}{\hspace*{-4cm}}\bf I.\,\, The leading term QCD predictions\end{center}

The leading contributions to the hard kernels for these amplitudes at large $s=W^2=(q_1+q_2)
^2$ and fixed c.m.s. angle $\theta$ were first calculated in \cite{BL2} for symmetric meson
wave functions, $\phi_M(x)=\phi_M(1-x)$,\, and later in \cite{Maurice} (BC in what follows)
for arbitrary wave functions. Two typical diagrams are shown in fig.1\,. The main features of
these cross sections are as follows (everywhere below we follow mainly the definite
predictions from BC in \cite{Maurice})\,.

The helicity amplitudes look as\,:
\bq
A^{(lead)}_{\lambda_1\lambda_2}(s,\theta)=\frac{64\pi^2}{9s}\,\alpha \,{\ov \alpha}_s \,f_K^2
\int_0^1 dx_s\phi_K(x_s)\int_0^1 dy_s\phi_K(y_s)\,T_{\lambda_1\lambda_2}(x_s,y_s,\theta).
\eq

The hard kernels $T_{\lambda_1\lambda_2}(x_s,y_s,\theta)$ are\,:
\bq
T_{++}=T_{--}=(e_s-e_u)^2\,\frac{1}{\sin^2\theta}\,\frac{A}{D}\,\,\,,
\eq
\bq
T_{+-}=T_{-+}=\frac{1}{D}\Biggl [ \frac{(e_s-e_u)^2}{\sin^2\theta}(1-A)+e_s e_u \frac
{A C}{A^2-B^2\cos^2\theta}+\frac{(e_s^2-e_u^2)}{2}(y_s-x_u)  \Biggr ]\,,\nonumber
\eq
where\,: $A=x_sy_u+x_uy_s\,,\,\,B=x_sy_u-x_uy_s\,,\,\,C=x_sx_u+y_sy_u\,,\,\, D=x_u x_s
y_u y_s\,,\, x_s+x_u=1\,,\, e_u=2/3,\,\,e_s=e_d=-1/3\,,\,f_M$ are the couplings\,:\, $f_{\pi}=
130\,{\rm MeV}\,,\,\, f_K=160\,{\rm MeV}\,,$ and $\phi_K(x_s)$ is the leading twist K-meson
wave function.
The above expressions are given for $K^+K^-$, other cases are obtained by evident replacements.
The pion wave function is symmetric,\, $\phi_{\pi}(x_u,x_d)=\phi_{\pi}(x_d,x_u)$, while
$\phi_K(x_s,x_u)- \phi_K(x_u,x_s)\neq 0$ due to $SU(3)$-symmetry breaking effects. Besides,
even the part of $\phi(x_s,x_u)$ symmetric in $x_u\leftrightarrow x_s$ differs from $\phi_
{\pi}(x_u,x_d)$, for the same reason.

The cross section is\,:
\bq
\frac{d\sigma(\gamma\gamma\ra K^+K^-)}{d\cos\theta}=\frac{1}{32\pi s}\,\frac{1}{4}\sum_{
\lambda_1 \lambda_2}\Bigl | A_{\lambda_1\lambda_2}\Bigr |^2\,.
\eq

\begin{figure}{\vspace*{-2.5cm}}
\centering{\hspace*{-2cm}}
\includegraphics[width=0.65\textwidth]{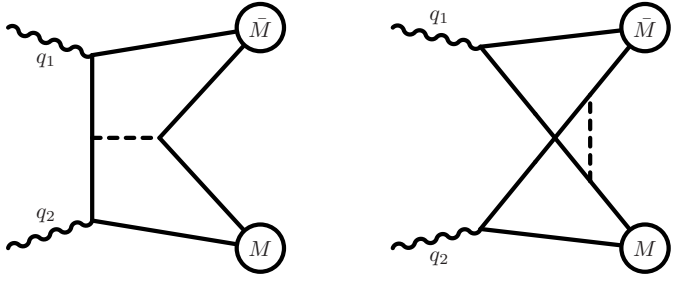}
\caption{Two typical Feynman diagrams for the leading term hard contributions to
$\gamma\gamma\ra {\ov M}M$\,, the broken line is the hard gluon exchange.}
\end{figure}

\vspace{0.3cm}
{\bf a)}\,\, The leading contribution to  $d\sigma(\pi^+\pi^-)$ can be written as\,:
\bq
\frac{s^3}{16\pi\alpha^2}\,\frac{d\sigma(\gamma\gamma\ra\pi^+\pi^-)}{d |\cos\theta |}
\equiv \frac{|\Phi^{(eff)}_{\pi}(s,\theta)|^2}{\sin ^4 \theta}
=\frac{|s F_{\pi}^{(lead)}(s)|^2}{\sin ^4 \theta}|1- \upsilon(\theta)|^2\,,
\eq
where
$F_{\pi}^{(lead)}(s)$ is the leading term of the pion form factor \cite{Ch1}\,:
\bq
|s F^{(lead)}_{\pi}(s)|= \frac{8\pi\,{\ov {\alpha}_s}}{9}\,\Bigl
|f_{\pi}\int_0^1 \frac{dx}{x}\,\phi_{\pi}(x,\,{\ov \mu})\Bigr |^2\,,
\eq
and $\upsilon(\theta)$ is due to the $\sim AC$ term in eq.(2). Below we will compare the
predictions of two frequently used models for $\phi_{\pi}(x): \phi^{(asy)}(x)=6x(1-x)$
and $\phi^{(CZ)}(x,\mu_o)=30x(1-x)(2x-1)^2,\,\,\mu_o=0.5\,{\rm GeV}$ \cite{Ch3}. While
the numerical value of $|s F^{(lead)}_{\pi}(s)|$ is highly sensitive to the form of
$\phi_{\pi}(x)$, the function $\upsilon(\theta)$ is nearly independent of $\theta$ at
$|\cos\theta|<0.6$ and, as emphasized in \cite{BL2}, is weakly sensitive to the form
of $\phi_{\pi}(x)$. For the above two very
different pion wave functions,\,\, $\upsilon(\theta)\simeq 0.12$.

The recent data from Belle  \cite{Belle1} for $(\pi^+\pi^-)$ and
$(K^+K^-)$ agree with $\sim 1/\sin^4\theta $ dependence at $W\geq
3\,GeV$, while the angular distribution is somewhat steeper at
lower energies. The energy dependence at $2.4\,{\rm GeV} < W < 4.1\, {\rm GeV}$
was fitted in \cite{Belle1} as: $\sigma_o(\pi^+\pi^-)=\int_0^{0.6}dc
(d\sigma/d|c|) \sim W^{-n}\,,\,\, \,n=(7.9\pm 0.4\pm 1.5)$ for
$(\pi^+\pi^-)$, and $n=(7.3\pm 0.3\pm 1.5)$ for $(K^+K^-)$. However, the
overall value $n\simeq 6 $ is also acceptable, see fig.2\,.

As for the absolute normalization, the $(\pi^+\pi^-)$ data are fitted \cite{Belle1} with\,:
$|\Phi_{\pi}^{(eff)} (s,\theta)|=(0.503\pm 0.007\pm 0.035)\,{\rm GeV}^2$.
\footnote{ \,\,\,Clearly, in addition to the leading terms $A^{(lead)}$ given by eqs.(1-5),
this experimental value includes also all loop and power corrections $\delta A$ to the $\gamma
\gamma\ra \pi^+\pi^- $ amplitudes $A=A^{(lead)}+\delta A$. These are different, of course,
from corrections $\delta F_{\pi}$ to the genuine pion form factor $F_{\pi}=F_{\pi}^{(lead)}+
\delta F_{\pi}$. So, the direct connection between the leading terms of $d\sigma(\pi^+\pi^-)$
and $|F_{\pi}|^2$ in eq.(4) does not hold on account of corrections.}
This value can be compared with \,: $0.88\cdot |s F_{\pi}^{(CZ)}(s)|\simeq 0.40\,{\rm GeV}^2$,
obtained by using $\phi_{\pi}(x)=\phi_{\pi}^{(CZ)}(x,\mu_o)$ in eq.(5). It
is seen that there is a reasonable agreement. At the same time, using $\phi_{\pi}(x)=
\phi^{(asy)}(x)$ one obtains much smaller value\,: $0.88\cdot |F_{\pi}^{(asy)}(s)|\simeq 0.13
\,{\rm GeV}^2$. So, for the pion wave function $\phi_{\pi}(x)$ close to $\phi^{(asy)}
(x)$ the leadind term calculation predicts the cross section which is $\simeq 15$ times
smaller than the data. It seems impossible that, at energies $s= 10-15\,{\rm GeV}^2$,
higher loop or power corrections can cure so large difference.
\footnote{\,\, A similar situation occurs in calculations of charmonium decays. ${\rm Br}
(\chi_o\ra \pi^+\pi^-)$ and ${\rm Br} (\chi_2\ra \pi^+\pi^-)$ calculated with $\phi_{\pi}(x)=
\phi^{(asy)}(x)$ are $\simeq 20-25$ times smaller than the data, while the use of $\phi_{\pi}
(x)=\phi^{(CZ)}(x,\mu_o)$ leads to values in a reasonable agreement with the data, see
\cite{Ch3},\cite{CZ}.}
\\

{\bf b)}\,\, The SU(3)-symmetry breaking, $d\sigma(K^+K^-)\neq d\sigma(\pi^+\pi^-),$
originates not only from different meson couplings, $f_K\neq f_{\pi}$, but also from symmetry
breaking effects in normalized meson wave functions, $\phi_{K}(x)\neq \phi_{\pi}(x)$. These
two effects tend to cancel each other, when using for the K-meson the wave function $\phi_
{K}(x_s,x_u)$ proposed in \cite{Ch5} (see \cite{CZ} for a review). So, instead of the naive
prediction $\simeq (f_K/f_{\pi})^4\simeq 2.3$ from \cite{BL2}, the prediction of BC for this
ratio is close to unity, and this agrees with the recent data from Belle \cite{Belle1}:
\bq
\frac{\sigma_o (\gamma\gamma\ra K^+K^-)}{\sigma_o (\gamma\gamma\ra \pi^+\pi^-)}=\left \{
\begin{array}[c]{ll}\displaystyle (f_K/f_{\pi})^4\simeq 2.3 \scriptstyle &
\begin{array}[c]{l}\hspace*{-4.0 cm}\rm{ Brodsky,\, Lepage} \,\, \cite{BL2} \end{array}
\\ & \\ \hline & \\ \simeq 1.06 & \begin{array}[c]{l} \hspace*{-4.7 cm} \rm{Benayoun,\,
Chernyak} \,\,\cite{Maurice}\end{array} \\ & \\ \hline & \\
(0.89\pm 0.04\pm 0.15)\hspace*{3.0 cm}  {\rm Belle} \,\,\, \cite{Belle1}
\end{array} \right. \nonumber
\eq

\begin{figure}{\hspace*{-2cm}}
\includegraphics[width=0.6\textwidth]{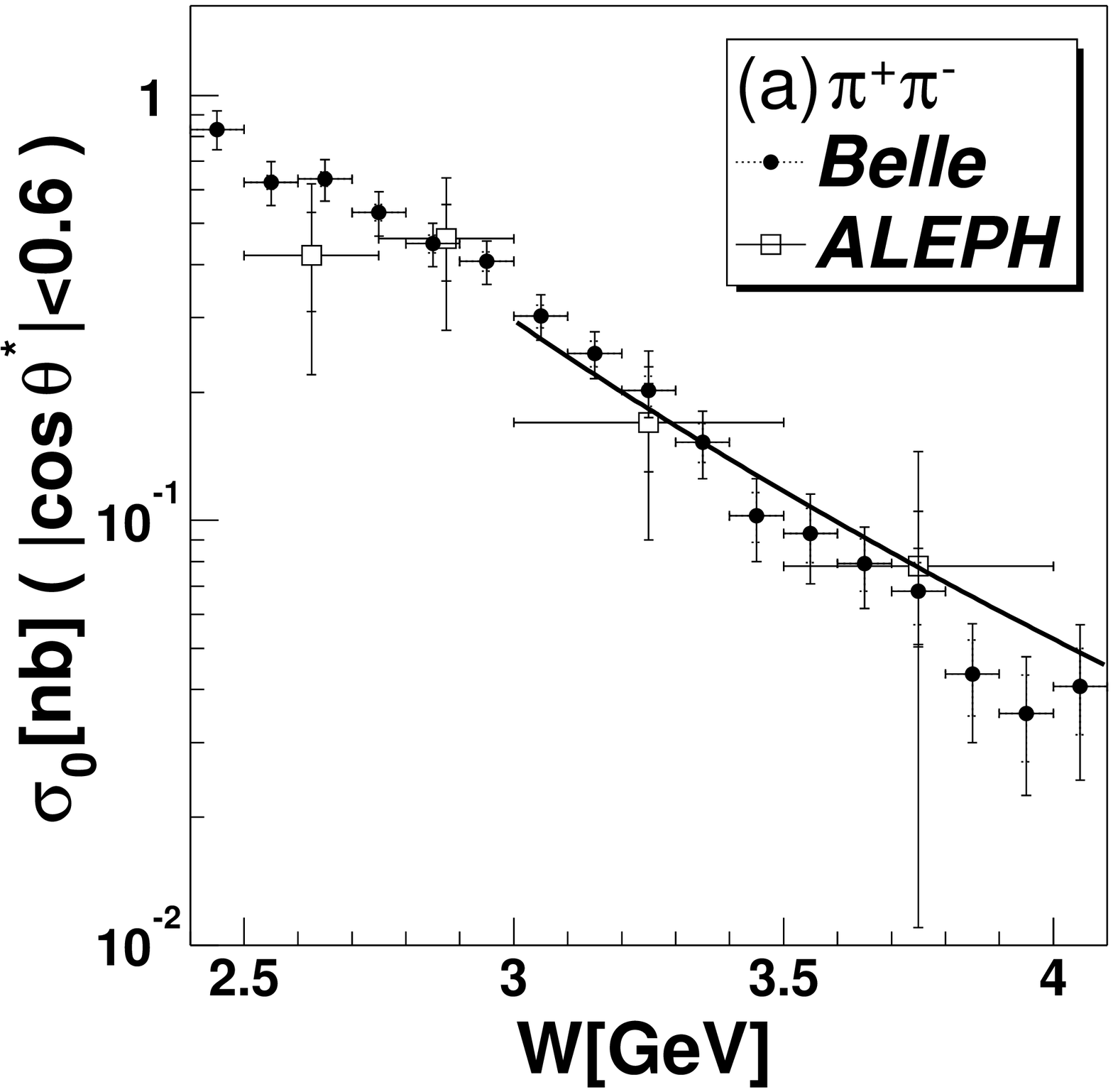}~
\includegraphics[width=0.6\textwidth]{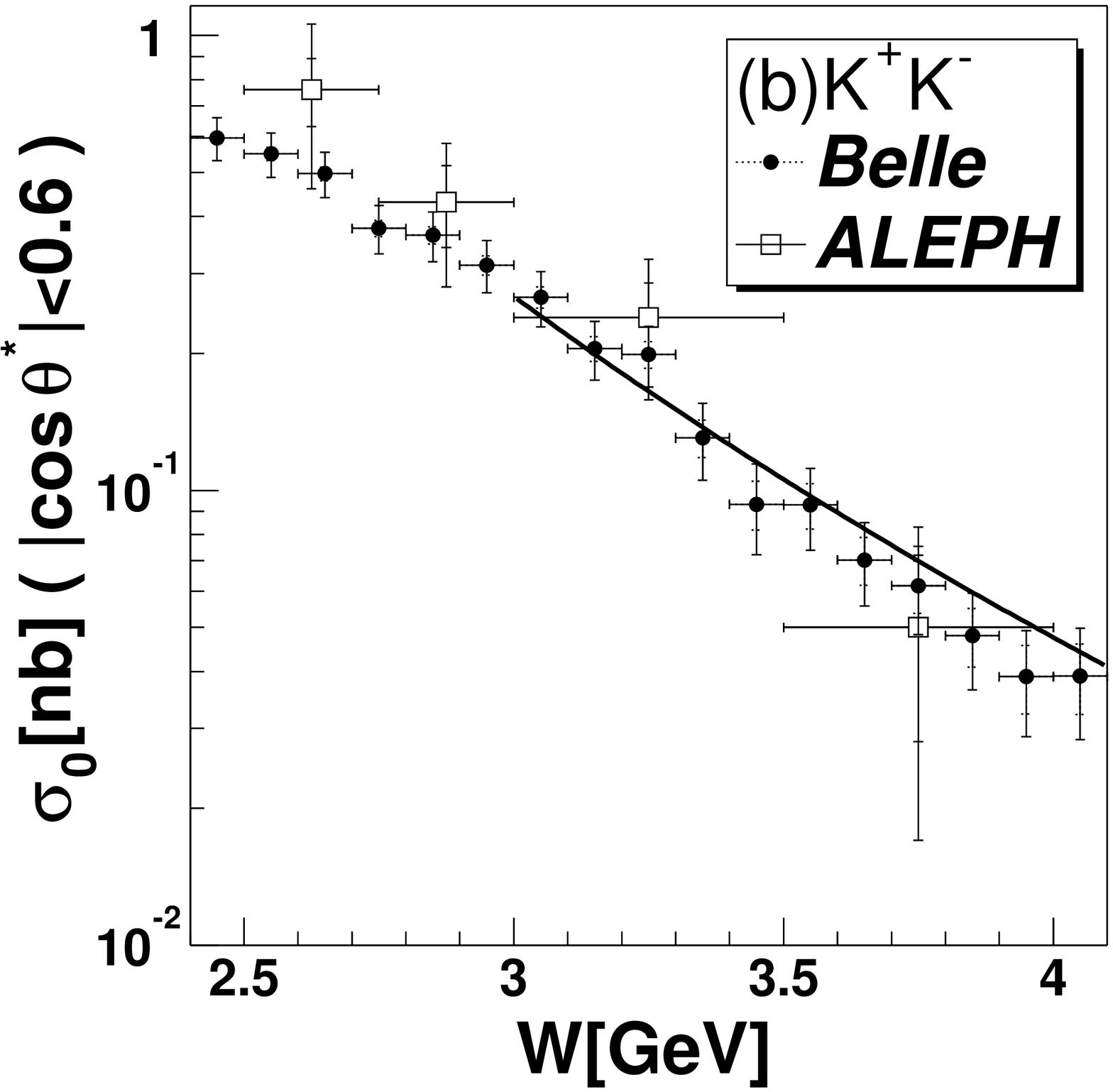}\\
\\
\centering
\includegraphics[width=0.9\textwidth]{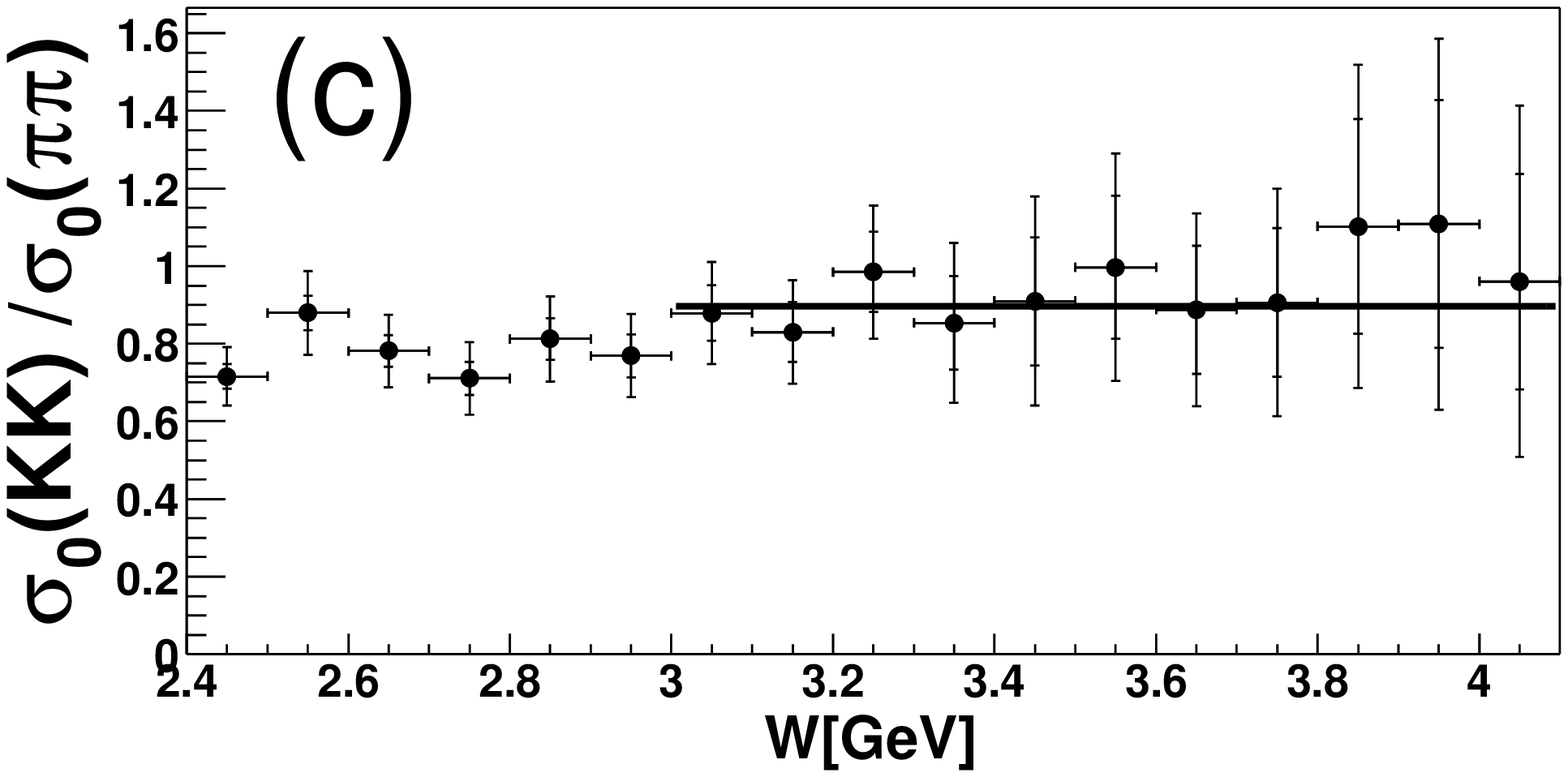}
\caption {\, The cross sections $ d\sigma/d\cos\theta $\,,\, integrated over the
c.m. angular region $|\cos\theta|<0.6\,$, together with a $ W^{-6}$ dependence line
\,\, \cite{Belle1}\,\,:\,\,  a) $\,\,\gamma \gamma\ra \pi^+\pi^-\,,\,\,$ b) $\,\,
\gamma\gamma\ra K^+K^-\,,$\,\, c) the cross section
ratio, the solid line is the result of the fit for the data above $3\,{\rm GeV} \,,\,\,$
the errors indicated by short ticks are statistical only.}
\end{figure}~

{\bf c)}\,\, The leading terms in the cross sections for neutral particles are much smaller
than for charged ones. For instance, it was obtained by BC that the ratio $d\sigma^{(lead)}
(\pi^o\pi^o)/d\sigma^{(lead)}(\pi^+\pi^-)$ varies from $\simeq 0.07$ at $\cos \theta =0$ to
$\simeq 0.04$ at $\cos\theta =0.6$, while the ratio\,: $({\ov K}^o K^o)^{(lead)}/(\pi^o
\pi^o)^{(lead)}\simeq 1.3\cdot(4/25)\simeq 0.21\,.$ So, one obtains for the
cross sections $\sigma_o^{(lead)}$ integrated over $0\leq |\cos\theta|\leq 0.6 $ for charged
particles and over $0 \leq \cos\theta \leq 0.6 $ for neutral ones\, :\,\, $\sigma_{o}^
{(lead)}(K_S K_S)/\sigma_{o}^{(lead)}(K^+K^-)\simeq 0.005\,.$

It is seen that the leading contribution to $\sigma_o(K_SK_S)$ is
very small. This implies that it is not yet dominant at
present energies $ W^2 < 16\,{\rm GeV}^2$. In other words, the amplitude
$A(\gamma\gamma\ra K_S K_S)= a(s,\theta)+b(s,\theta)$ is dominated
by the non-leading term $b(s,\theta)\sim g(\theta)/s^2$, while the
formally leading term $a(s,\theta)\sim C_o f_{BC}(\theta)/s$ has so small
coefficient $C_o$ that $|b(s,\theta)| > |a(s,\theta)|$ at, say, $
W^2 < 12\,{\rm GeV}^2$. Therefore, it has no meaning to compare the leading term
prediction of BC (i.e. $d\sigma(K_S K_S)/d \cos\theta\sim
|a(s,\theta)|^2/s \sim |f_{BC}(\theta)|^2/W^{6} $ at $s\ra \infty$)
for the energy and angular dependence of $d\sigma(K_S K_S)$ with the
recent data from Belle \cite{Belle2}. Really, the only QCD
prediction for $6\,{\rm GeV}^2 < W^2 < 12\,{\rm GeV}^2$ is the energy
dependence: $d\sigma(K_S K_S)/d\cos\theta \sim |b(s,\theta)|^2/s\sim
|g(\theta)|^2/W^{10}$, while the angular dependence $|g(\theta)|^2$
and the absolute normalization are unknown. This energy dependence
agrees with the recent results from  Belle \cite{Belle2}, see fig.4\,.

\begin{center}{\hspace*{-4cm}}{\bf II. \,\, The handbag model predictions}
\end{center}

{\bf The hand-bag model} \cite{DKV} (DKV in what follows) is a part of a general ideology
which claims that present day energies are insufficient for the leading terms QCD to be the
main ones. Instead,  the soft nonperturbative contributions are supposed to dominate the
amplitudes. The handbag model represents applications of this ideology to description of
$d\sigma(\gamma \gamma\ra {\ov M}M)$. It assumes that the above described hard contributions
really dominate at very high energies only, while the main contributions at
present energies originate from the fig.3a diagram. Here, two
photons interact with the same quark only, and these "active" ${\ov
q}q$-quarks carry nearly the whole meson momenta, while the
additional "passive" ${\ov q^\prime}q^ \prime $ quarks are "wee
partons" which are picked out from the vacuum by soft
non-perturbative interactions. It was obtained by DKV that the
angular dependence of amplitudes is $\sim 1/\sin^2\theta$, while the
energy dependence is not predicted and is described by some soft
form factors $ R_{M}(s), $ which are then fitted to the data.
Because the "passive" quarks are picked out from the vacuum by soft
forces, these soft form factors should be power suppressed at large $s\,:
R_{M}(s) \leq 1/s^2\,,\, $ in comparison with the leading meson form
factors, $F_{M}(s)\sim 1/s\,$.
\begin{figure}{\vspace*{-2.5cm}}
\centering{\hspace{-2cm}}
\includegraphics[width=0.9\textwidth]{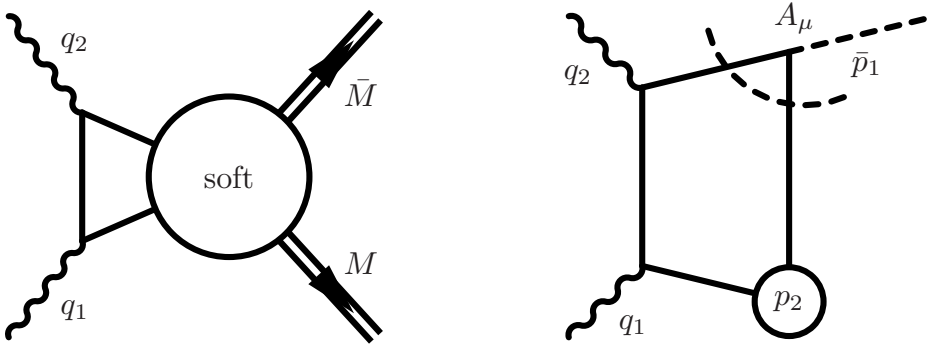}
\caption{a) the overall picture of the handbag contribution,\, b) the lowest
order  Feynman diagram for the light cone sum rule}
\end{figure}

{\bf a)}\,\, As for the flavor and charge dependence, the predictions of the
handbag model look as follows:
\bq \frac{ d\sigma(\gamma\gamma\ra \pi^o\pi^o)/d\cos\theta} {d\sigma(\gamma\gamma\ra
\pi^+\pi^-)d|\cos\theta|}\equiv \frac{(\pi^o\pi^o)}{(\pi^+\pi^-)}=\frac{1}{2}\Biggl |
\frac{(e_u^2\,R_{u\ra u}(s)+e_d^2\, R_{d\ra d}(s))/2}{e_u^2\,R_{u\ra
d}(s)+e_d^2\,R_{d\ra u}(s)}\Biggr |^2 =\frac{1}{2}\,.\nonumber
\eq
Here, $e_u=2/3,\,\, e_d=e_s=-1/3$ are the quark charges, while the
form factor $R_{u\ra d} (s)$ corresponds to the active $u$-quark and
passive $d$-quark, etc. The isotopic symmetry implies that\,:
$R_{u\ra u}=R_{u\ra d}=R_{d\ra u}=R_{d\ra d}$. Unfortunately, the cross section
$d\sigma(\gamma\gamma\ra \pi^o\pi^o)/d\cos\theta$ is not measured up to now.

As for the $SU(3)$-symmetry breaking effects, it seems clear that it
is harder for soft interactions with the scale $\sim \Lambda_{QCD}$
to pick out from the vacuum a heavier ${\ov s}s$-pair, than the
light ${\ov u}u$-\, or ${\ov d}d$-pairs.
\footnote{\,\,\, The effect due to $m_s\neq 0$ of the hard quark propagating
between two photons in fig.3 is small and can be neglected , see the Appendix. }
So\,:\,\, $ R_{u\ra s}(s)/R_{s\ra u}(s)= (1-2\Delta)\,,\,\, \Delta > 0$.\,\,
(\,The same inequality $\Delta > 0$ follows from the fact that the heavier
s-quark carries the larger mean fraction of the K-meson momentum, $\langle x_s
\rangle \,>\,0.5 $ ). Therefore, the handbag model predicts:
\bq
\frac{(K_S K_S)}{(K^+K^-)}=\frac{1}{2}\Biggl |\frac{e^2_s\,R_{s\ra d}(s)+e_d^2\,R_{d\ra
s}(s)} {e_u^2\,R_{u\ra s}(s)+e_s^2\,R_{s\ra u}(s)}  \Biggr
|^2=\frac{2}{25}\Biggl |\frac {1-\Delta}{1-\frac{8}{5}\Delta}\Biggr
|^2 > 0.08\,.
\eq
\\

{\bf b)}\,\, To obtain more definite predictions, in comparison with DKV \cite{DKV}, we
have directly calculated the soft contributions of the handbag model to $\gamma\gamma\ra
\pi\pi,\,KK$.

Below we present in a short form the explicit calculation of the
handbag diagram contribution to the amplitude $A(\gamma\gamma\ra \pi^+\pi^-)$,
the results for $K^+K^-$ and $K_S K_S$ are then obtained
by trivial replacements.

To deal with such soft contributions we use the method of
"the light cone sum rules" \cite{Braun},\,\cite{NN}. One pion
with the momentum $p_1\,,\,p_1^2=m_{\pi}^2\simeq 0$ is replaced by
the interpolating current $A_{\mu}={\ov u}\gamma_{\mu}\gamma_5 d$
with the momentum ${\ov p}_1 \,,\,{\ov p}_1^2\neq 0$, while the
second pion $\pi(p_2)$ stays intact, see fig.3b\,. The amplitude ${\ov A}$ is
calculated which is\,:\,\,${\ov A}=(q_1-q_2)_\mu\,\langle \pi(p_2)|
A_{\mu}(0)|\gamma(q_1),\,\gamma(q_2)\rangle\,$. Its discontinuity in
${\ov p}_1^2$ has the form\,: \,$(1/\pi)\,\Delta  {\ov
A}=f_{\pi}\,(q_1-q_2){\ov p}_1\,\delta({\ov p}_1^2)\,A(\gamma
\gamma\ra\pi^+\pi^-)+\theta({\ov p}_1^2-s_{\pi}){\ov A}^{(cont)},$\,
where $\,(q_1-q_2) {\ov p}_1=(u-t)/2\,,$ and $s_{\pi}\simeq
0.7\,{\rm GeV}^2$ is the pion duality interval. The sum rule is written
for the quantity
\bq
\Bigl [f_{\pi}\,(u-t)/2\Bigr
]^{-1}\int_0^{s_{\pi}}d{\ov p}_{1}^2\,\exp\{-{\ov p}_1^2/M^2\}\,
(1/\pi)\,\Delta {\ov A}=A(\gamma\gamma\ra \pi^+\pi^-).\nonumber
\eq

When calculating this quantity from the fig.3b diagrams, one
proceeds first in the Euclidean region ${\ov s}=-s=-(q_1+q_2)^2 >0$,
and the final result is then continued analytically into the
physical region $s > 0$.
\footnote{\,\, This approach was used in \cite{NN} for the calculation of the B-meson decays
into two baryons.  The kinematics is Minkowskian in these decays, as in
$\gamma\gamma\ra {\pi^+\pi^-}$.}
The calculation here is simple and straightforward, we note only that the leading
twist wave function $\phi_{\pi}(x)$ is used for $\pi(p_2)$. One obtains \,:
\bq
A_{handbag}(\gamma\gamma\ra\pi^+\pi^-)=16\pi\alpha\,\Biggl(e_u^2\,R_{u\ra d}(s)+e_d^2\,
R_{d\ra u}(s)\Biggr )\,(1+
\epsilon_{\mu}^{(1)}\epsilon_{\mu}^{(2)}),
\eq
\bq
R_{u\ra d}(s)=\frac{\omega}{2}\int_o^{1}\frac{d z\,\phi_{\pi}(x_u=1-\omega z, x_d=\omega z)}
{(1-\omega z)}\,
\exp \Bigl \{-\,\frac{z}{(1-\omega z)(1+s_{\pi}/{\ov
s})}\,\frac{s_{\pi}}{M_o^2}\Bigr \}\,,\nonumber
\eq
where $\omega=s_{\pi}/({\ov s}+s_{\pi})$,\,\, $\epsilon^{(1)}$ and
$\epsilon^{(2)}$ are the polarization vectors of two photons,
$M_o^2$ is the optimal value of $M^2$ (typically, $M_o^2\simeq
m_{\rho}^2\simeq 0.6\, {\rm GeV}^2$), and $R_{d\ra u}(s)$ is obtained from $R_{u\ra d}(s)$
by the replacement: $\phi_{\pi}(x_u=1-\omega z, x_d=\omega z)\ra \phi_{\pi}(x_d=1-\omega z,
x_u=\omega z)$.

It is seen from eq.(7) that the handbag amplitude is independent of
the scattering angle $\theta$, in contradiction with the DKV result \cite{DKV}
about $\sim 1/\sin^2\theta$ dependence. As for the energy dependence, using
$\phi_{\pi}(x)\ra C_{\pi}x$ at $x\ll 1$, where $C_{\pi}=\rm const$,
one has: $A_{handbag}(\gamma\gamma\ra\pi^+\pi^-)\sim R(s)\sim
(s_{\pi}/s)^2$, as expected.

For comparison with eq.(4), let us define the effective "handbag
form factor" $\Phi^{hb}_{\pi}(s)$ for $\pi^+\pi^-$ as:
\bq
\frac{s^3}{16\pi\alpha^2}\,\frac{d\sigma^{hb}(\gamma\gamma\ra \pi^+\pi^-)}{d |\cos\theta |}
\equiv | \Phi^{hb}_{\pi}(s)|^2\,,\quad \Bigl |\Phi^{hb}_{\pi}(s)\Bigr |= \Bigl |
\frac{(e_u^2+e_d^2)}{\sqrt 2}s\,R_{u\ra d}(s)\,\Bigr |\,.
\eq
Its numerical value depends strongly on the form of $\phi_{\pi}(x)$, see eq.(7). For
$\phi_{\pi}(x)=\phi^{(asy)}(x)=6x(1-x),\,\, |\Phi^{hb}_{(asy)}(s)|\,$ is\,:\,\,$ 0.029\,
{\rm GeV}^2\,\,\,\rm{at}\,\,\,s=6\,{\rm GeV}^2,\\ 0.016\,{\rm GeV}^2\,\,\, \rm{at}\,\,\,
s=10\,{\rm GeV}^2$, and $0.009\,{\rm GeV}^2\,\,\, \rm{at}\,\,\, s=16\,{\rm GeV}^2$. The
corresponding values for $\phi_{\pi}^{(CZ)}(x,\mu_o)=30x(1-x)(1-2x)^2$ are respectively\,
:\, $0.19\,{\rm GeV}^2,\,\,0.09\,{\rm GeV}^2$ and $0.05\,{\rm GeV}^2$. It is seen that, even
for $\phi_{\pi}^{(CZ)}(x,\mu_o)$,\,\, $|\Phi ^{hb}_{\pi}(s)|$ is small in comparison with
the experimental value\,:\,$|\Phi_{\pi}^{(eff)}(s,\theta)|\simeq\rm{const}=0.50\, {\rm GeV}
^2$, see eq.(4), and, besides, decreases strongly with increasing energy. So, even at available
energies, these soft form factors $R_i(s)$ "honestly" decay with the non-leading power,
$R_i(s)\sim 1/s^2$, and do not imitate the leading behavior $\sim 1/s.$ \\

Neglecting terms $O(M_K^2/s)$, one obtains for $A_{handbag}(\gamma\gamma\ra K^+K^-)$ the
same eq.(7), with $\phi_{\pi}(x_u,x_d)\ra \phi_{K}(x_u,x_s)\exp \{ m_K^2/M^2 \}$ and
$s_{\pi}\ra s_K $. Using for $\phi_{K}(x_u,x_s)$ the wave function $\phi_{K}^{(CZ)}
(x_s,x_u,\mu_o)$ proposed in \cite{Ch5}, one then obtains:
\bq
\Delta =\frac{1}{2}\, \Biggl (\,\, 1- \frac{R_{u\ra s}(s)}{R_{s\ra u}(s)}
\,\,\Biggr )\simeq 0.25\,.
\eq

Besides, one can calculate also the $SU(3)$-symmetry breaking
effects in $(\pi^+\pi^-)/ ( K^+K^-)$. For instance, with $M_o^2\simeq
M^2_{K^{*}}\simeq 0.8\,{\rm GeV}^2,\, s_K\simeq 0.93\, {\rm GeV}^2$, one obtains
the prediction of the handbag model\,:
\bq
\frac{(K^+K^-)_{hb}}{(\pi^+\pi^-)_{hb}}=\Bigl | \frac{R_{s\ra u}(s)}{R_{d\ra
u}(s)}\Bigl (1-\frac{8}{5} \Delta \Bigr ) \Bigr |^2 \simeq \bigl |
0.86 \Bigr |^2\simeq 0.73\,.
\eq

On the whole, the above described predictions of the handbag model for $(\pi^+\pi^-)$
and $(K^+K^-)$ disagree with the data \cite{Belle2} both in the energy dependence and
the angular dependence. Besides, even for the wide wave functions $\phi_{\pi}^{(CZ)}(x,\mu_o)$
\cite{Ch3} and $\phi_{K}^{(CZ)}(x,\mu_o)$ \cite{Ch5}, the predictions of the handbag model
for the absolute values of $\sigma_o(\pi^+\pi^-)$ and $\sigma_o(K^+K^-)$ are too small.
For instance, one obtains from eqs.(7,8) that at $s=10\,{\rm GeV}^2$ and $\phi_{\pi}(x)=
\phi_{\pi}^{(CZ)}(x,\mu_o)$,\,\, $\sigma_o^{(hb)}(\pi^+\pi^-)$ is $\simeq 40-50$ times
smaller than the experimental value.

As for the handbag model predictions for $(K_SK_S)$, one has for the ratio in eq.(6) (\, see
eq.(9))\,:
\bq
\frac{(K_SK_S)_{hb}}{(K^+K^-)_{hb}}= \frac{2}{25}\,\Biggl | \frac
{1-\Delta}{1-\frac{8}{5}\Delta}\Biggr |^2\simeq 0.08\,
\Bigl |\, 1.25\, \Bigr |^2\simeq 0.12\,.
\eq
%\vspace*{3mm}

{\bf c)}\,\, The cross section $d\sigma (K_SK_S)/d\cos\theta $ has been measured recently by
the Belle collaboration \cite{Belle2}. The energy dependence at $2.4\,{\rm GeV} < W < 4.0\,
{\rm GeV}$ was found to be\,:\,$\sigma_o(K_S K_S)\sim W^{-k}\,,\,k=(9.93\pm 0.44)$,\,
see fig.4\,. This agrees with the qualitative QCD prediction, see above, that the formally
leading ( at sufficiently large $s$) term $a(s,\theta)\sim C_o f_{BC}(\theta)/s$\, in the
amplitude $A(\gamma\gamma\ra K_SK_S)=a(s,\theta)+b(s,\theta)$, has a very small numerical
coefficient $C_o$, so that the non-leading term $b(s,\theta)\sim g(\theta)/s^2$  is really
larger at present energies. This is seen also from the absolute value of $\sigma_o(K_SK_S)$
measured by Belle \cite{Belle2}, see fig.4\,. Even at the highest energy $W\simeq 3.8\,
{\rm GeV}$, it is still above the value of $\sigma_o^{(lead)}(K_SK_S)$, obtained
by BC for the leading at $s\ra \infty$ term. The same can be seen, of course, from $\sigma_o
(K_SK_S)/\sigma_o(K^+K^-)$. While the value of $\sigma_o^{(lead)}(K^+K^-)$
predicted by BC is in a reasonable agreement with the data at present energies \cite{Belle1},
the prediction of BC for the ratio of the leading at $s\ra \infty$ contributions\,: $
\sigma_o^{(lead)}(K_SK_S)/\sigma_o^{(lead)}(K^+K^-)\simeq 0.005$, is still below its
values at present energies, see figs.(2b,\,4)\,. In other words, for the formally leading term
$|a(s,\theta)|^2$ to be really dominant in $d\sigma(K_SK_S)/d\cos\theta$, the energy $W$ has
to be increased. Only then one will see the behavior $\sigma_o^{(lead)}(K_SK_S)\sim 1/W^6$.

Because the leading at large $s$ term in $d\sigma (K_SK_S)/d\cos\theta $ has so small
coefficient, this process is the ideal place for the handbag model to be applicable. As it
is, the handbag diagram calculated above contributes to the non-leading term $b(s,\theta)\,
:\,\,b_{hb}(s,\theta)={\rm const}/s^2$. This agrees with the data in the energy dependence,
but predicts the flat angular dependence $d\sigma(K_S K_S)/d\cos \theta \sim {\rm const}/
W^{10}$, while the data prefer $d\sigma(K_S K_S)/d\cos \theta \sim 1/(W^{10}\sin^4\theta)$\,
\cite{Belle2}. As for the absolute value of $\sigma_{o}^{(hb)}(K_SK_S)$, one obtains from
the above that at $s=10\,{\rm GeV}^2$ and even for the wide $\phi_{\pi}(x)=\phi_{\pi}^{CZ}(x,
\mu_o)$ it is $\simeq 10$ times smaller than the experiment, and more than two orders smaller
for $\phi_{\pi}(x)=\phi^{(asy)}(x)$.

So, even in this process, the handbag contribution does not dominate.
\footnote{\,\, One has to remember that there are also other contributions to the
non-leading term $b(s,\theta)$ with the same energy dependence $\sim 1/s^2$ (and
unknown angular dependence and absolute normalization), in addition to the handbag one.}

\begin{figure}{\vspace*{-3.3 cm}}~
\centering{\hspace*{-1.5 cm}}
\includegraphics[width=0.7\textwidth]{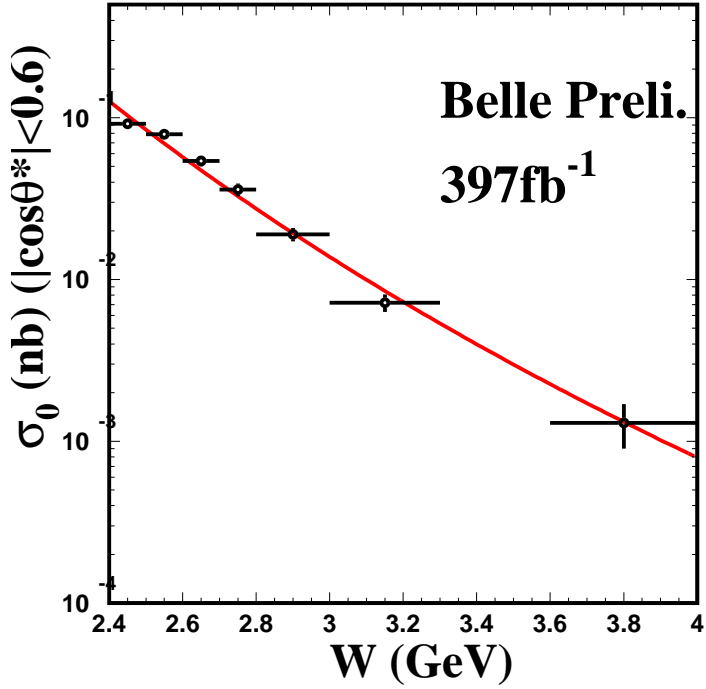}~
\caption{ The measured energy dependence of $\sigma_o(K_SK_S)$\,\, \cite{Belle2}.
The solid line is $\,\sim W^{-k},\,\, k\simeq 10$.}
\end{figure}~

The measured ratio $(K_S K_S) /(K^+K^-)$ decreases strongly with increasing energy, and
becomes smaller than the lower bound $0.080$ in eq.(6) at $W > 2.8\,{\rm GeV}$\,
\cite{Belle2},\, see figs.(2b,\,4). This is also in contradiction with the handbag model
predictions, see eq.(6) and eq.(11).\footnote{\,\, The ratio $\sigma_o(K_SK_S)/\sigma_o
(K^+K^-)$ of measured cross sections \cite{Belle1},\,\cite{Belle2} is\,: $\simeq 0.030$ at $W
=3.2\,{\rm GeV}$,\, and $\simeq 0.015$ at $W=3.8\,{\rm GeV}\,$, see figs.(2b,\,4)\,.}

\begin{center}{\bf Conclusions} \end{center}

Our conclusion is that the leading term QCD predictions are in  reasonable agreement
with the recent data from Belle, but only for the wide pion and kaon wave functions,
like $\phi_{\pi,\,K}^{(CZ)}(x)$. At the same time, the handbag model contradicts these data.
\newpage
\begin{center}{\bf Acknowlegements}\end{center}

I am grateful to A.E. Bondar and S.I. Eidelman for useful discussions about
experimental results.

\begin{center}{\bf Appendix}\end{center}

Besides the above described $SU(3)$-symmetry breaking effects originating from the
K-meson wave function, there is an additional handbag contribution to $K^+K^-$ from
$m_s\neq 0$ in the hard quark propagator in fig.3b\,. It seems at
the first sight that, relative to $A_{hb}(\pi^+\pi^-)$, it is additionally
suppressed by the factor $O(M_K^2/s)$ and can be neglected in
comparison with the corrections $O(M_K^2/s_K)$ from the $SU(3)$-symmetry
breaking effects in the K-meson wave function $\phi_K(x)$ in
eq.(7). Really, it gives also $O(M_K^2/s_K)$ correction. The reason
is that this contribution includes the twist-tree wave function
$\phi_{P}(x)$, originating from $\langle K|{\ov u}(z)\gamma_5
s(0)|0\rangle $. It behaves as $\phi_{P}(x)\sim 1$ at $x\ll 1$
(compare with $\phi_{K}(x)\sim x \sim (s_K/s)$ in eq.(7)), so that
this correction is also $\sim (M_K^2/s)(s/s_K)\sim
(M_K^2/s_K)$. One obtains for this additional contribution $\delta
A_{handbag}\,\,:$
\bq
\delta A_{handbag}(\gamma\gamma\ra K^+K^-)=16\pi\alpha\, e_s^2\, \frac{2 M_K^2}{s}\,\frac{
\epsilon_{\mu}^{(1)}\epsilon_{\mu}^{(2)}}{\sin^2\theta}\,\,I_P\,,\nonumber
\eq
\bq
I_P=\omega\int_o^{1}\frac{d z\,\phi_{P}(\omega z)}{(1-\omega z)^2}\,\exp \Bigl \{-\,\frac{z}
{(1-\omega z)(1+s_K/{\ov s})}\,\frac{s_K}{M_o^2}\Bigr \}\,.
\eq

It is seen from comparison of eq.(7) and eq.(12) that the helicity
structures are different, so that these two contributions do not
interfere in the cross section. As for the numerical value of
$\delta A_{handbag}$, using $\phi_{P}(x)=1$ and $e_s^2=1/9$, one
obtains that $|\delta A_{handbag}|^2$ is very small (about $100$
times smaller than even $|A_{hb}(\gamma\gamma\ra\pi^+\pi^-)|^2$ from
eq.(7) with $\phi_{\pi}(x)=\phi^{(asy)}(x))$, and can be neglected.\\


\newpage


\begin{thebibliography}{99}
\bibitem{BL2}
S.J. Brodsky, G.P. Lepage, Phys. Rev. {\bf D24} (1981) 1808
\bibitem{Maurice}
M. Benayoun, V.L. Chernyak, Nucl. Phys. {\bf B329} (1990) 285
\bibitem{Ch1}
V.L. Chernyak, A.R. Zhitnitsky, JETP Lett. {\bf 25} (1977) 510
\bibitem{Ch3}
V.L. Chernyak, A.R. Zhitnitsky, Nucl. Phys. {\bf B201} (1982) 492 ,\\
Erratum: {\it ibid} {\bf B214} (1983) 547
\bibitem{Belle1}
H. Nakazawa et. al. (Belle Coll.), Phys. Lett. {\bf B615}
(2005) 39 ;\\ hep-ex/0412058
\bibitem{Ch5}
V.L. Chernyak, A.R. Zhitnitsky, I.R. Zhitnitsky, Nucl. Phys. {\bf
B204} (1982) 477 ,\quad Yad. Fiz. {\bf 38} (1983) 1277
\bibitem{CZ}
V.L. Chernyak, A.R. Zhitnitsky, Phys. Rep. {\bf 112} (1984) 173
\bibitem{Belle2}
W.T. Chen (Belle Coll.), "Measurements of $K_S K_S$
production and charmonium studies in two-photon processes at Belle",
talk at the Int. Conf. "Photon-2005", September 3, 2005, Warsaw
and Kazimierz, Poland
\bibitem{DKV}
M. Diehl, P. Kroll, C. Vogt, Phys. Lett. {\bf B532} (2002) 99 ;\\
hep-ph/0112274
\bibitem{Braun}
I.I. Balitsky, V.M. Braun, A.V. Kolesnichenko, Nucl. Phys. {\bf B312} (1989) 509;
V.M. Braun, I.E. Filyanov, Z. Phys. {\bf C44} (1989) 157
\bibitem{NN}
V.L. Chernyak, I.R. Zhitnitsky, Nucl. Phys. {\bf B345} (1990) 137




\end{thebibliography}
\end{document}